\documentstyle[ws-procs975x65]{article}

\def\Journal#1#2#3#4{{#1} {\bf #2}, #3 (#4)}

\def\PR{\em Phys. Rept.}

\def\NPB{{\em Nucl. Phys.} B}
\def\NPA{{\em Nucl. Phys.} A}
\def\PLB{{\em Phys. Lett.}  B}
\def\PL{\em Phys. Lett.}
\def\PRL{\em Phys. Rev. Lett.}
\def\PRA{{\em Phys. Rev.} A}
\def\PRC{{\em Phys. Rev.} C}
\def\PRD{{\em Phys. Rev.} D}

\def\JETP{\em JETP Lett.}
\def\JPA{{\em J. Phys.}A} 

\def\be{\begin{equation}}
\def\ee{\end{equation}}
\def\bea{\begin{eqnarray}}
\def\eea{\end{eqnarray}}

\begin{document}

\title{SHAPE-INVARIANCE AND MANY-BODY PHYSICS} 

\author{A.~B. BALANTEKIN}

\address{University of Wisconsin, Department of Physics \\
Madison, WI  53706,  USA \\ 
E-mail: baha@nucth.physics.wisc.edu}

\maketitle

\abstracts{Recent developments in the study of shape-invariant
  Hamiltonians are briefly summarized. Relations between certain
  exactly solvable problems in many-body physics and shape-invariance
  are explored. Connection between Gaudin algebras and
  supersymmetric quantum mechanics is pointed out.} 

\section{Introduction}\label{1}

\indent
Supersymmetric Quantum Mechanics (SSM) is the name given to the study
of particular pairs of
Hamiltonians \cite{Witten:nf,Cooper:1994eh}. SSM can be motivated by
considering the ground ground state wavefunction,
$\psi_0(x)$, for a one-dimensional bound system. Since $\psi_0(x)$ has
no nodes it can be written as
\begin{equation}
  \label{eq:1}
  \psi_0(x) = \exp \left( - \frac{\sqrt{2m}}{\hbar} \int W(x) dx
  \right) ,
\end{equation}
where the function $W(x)$ is related to the potential energy of the
system. Introducing the operators 
\begin{eqnarray}
  \label{2}
  \hat{A} &=& W(\hat{x}) + \frac{i}{\sqrt{2m}} \hat{p}, \nonumber \\
  \hat{A}^{\dagger} &=& W(\hat{x}) - \frac{i}{\sqrt{2m}} \hat{p} ,
\end{eqnarray}
one can write the Hamiltonian of the system as
\begin{equation}
  \label{eq:2}
  \hat{H} - E_0 = \hat{A}^{\dagger}\hat{A},
\end{equation}
where $E_0$ is the ground state energy. The ground state
wavefunction satisfies the condition
\begin{equation}
  \label{eq:3}
  \hat{A}| \psi_0 \rangle = 0.
\end{equation}
It is straightforward to show that the supersymmetric partner
potentials 
\begin{eqnarray}
  \label{5}
  \hat{H}_1 &=& \hat{A}^{\dagger} \hat{A} \nonumber \\
  \hat{H}_2 &=& \hat{A} \hat{A}^{\dagger} 
\end{eqnarray}
have the same energy spectra except the ground state of $\hat{H}_1$,
the energy of which is zero. Potentials corresponding to these
Hamiltonians are
\begin{eqnarray}
  \label{6}
  V_1(x) &=& [W(x)]^2 - \frac{\hbar}{\sqrt{2m}} \frac{dW}{dx}
  \nonumber 
  \\
  V_2(x) &=& [W(x)]^2 + \frac{\hbar}{\sqrt{2m}} \frac{dW}{dx} .
\end{eqnarray}
The partner potentials in Eq. (\ref{6}) are called 
shape-invariant \cite{Gendenshtein:vs} if they can be obtained from
one another by changing their parameters: 
\begin{equation}
  \label{7}
  V_2(x;a_1) = V_1 (x;a_2) + R(a_1),
\end{equation}
where $a_2$ is a function of $a_1$, and the remainder $R(a_1)$ is
independent of $x$. Eq. (\ref{7}) is equivalent to the operator
relation 
\begin{equation}
  \label{eq:8}
  \hat{A}(a_1)\hat{A}^{\dagger}(a_1) = \hat{A}^{\dagger}(a_2)
  \hat{A}(a_2) + R(a_1) .
\end{equation}

\subsection{Algebraic Approach}

Shape-invariance problem was formulated in algebraic terms in
Ref. [4]. In this formulation one introduces an
operator which transforms the parameters of the potential:
\begin{equation}
  \label{eq:9}
\hat{T}(a_1) { O}(a_1) \hat{T}^{-1}(a_1) = { O}(a_2).
\end{equation}
Defining the operators 
\begin{eqnarray}
  \label{9a}
  \hat{B}_+ &=& \hat{A}^{\dagger}(a_1) \hat{T} (a_1) \nonumber \\
 \hat{B}_- = \hat{B}_+^{\dagger} &=& \hat{T}^{\dagger}(a_1) \hat{A}
 (a_1) 
\end{eqnarray}
one can show that the Hamiltonian can be written as
\begin{equation}
  \label{eq:10}
\hat{H} - E_0 = \hat{A}^{\dagger}\hat{A} = \hat{B}_+ \hat{B}_- .
\end{equation}
Using the definitions given in Eq. (\ref{9a}), the shape-invariance
condition of Eq. (\ref{eq:8}) takes the form
\begin{equation}
  \label{eq:11}
[ \hat{B}_- , \hat{B}_+ ] = R(a_0),
\end{equation}
where $R(a_0)$ is defined via
\begin{equation}
  \label{eq:12}
R(a_n) = \hat{T}(a_1) R(a_{n-1})\hat{T}^{\dagger}(a_1).
\end{equation}
In terms of these new operators Eq. (\ref{eq:3}) takes the form 
\begin{equation}
  \label{eq:13}
\hat{B}_- | \psi_0 \rangle = 0 ,
\end{equation}
i.e. the ground state is annihilated by the lowering operator
$\hat{B}_-$. 

One can easily establish the commutation
relations \cite{Balantekin:1997mg} 
\begin{equation}
  \label{eq:14}
[\hat{H}, \hat{B}_+^n ] = (R(a_1)+R(a_2)+ \cdot \cdot + R(a_n))
  \hat{B}_+^n
\end{equation}
\begin{equation}
  \label{eq:15}
[\hat{H}, \hat{B}_-^n ] = - \hat{B}_-^n (R(a_1)+R(a_2)+ \cdot \cdot
  + R(a_n))\,.
\end{equation}
i.e., $\hat{B}_+^n |\psi_0 \rangle$ is an eigenstate of the
Hamiltonian
with
the eigenvalue $R(a_1)+R(a_2)+ \cdot \cdot + R(a_n)$.
The normalized eigenstate is
\begin{equation}
  \label{eq:16}
|\psi_n \rangle = \frac{1}{\sqrt{R(a_1)+ \cdot \cdot +
R(a_n)}} \hat{B}_+ \cdot \cdot \frac{1}{\sqrt{R(a_1)+
R(a_2)}} \hat{B}_+  \frac{1}{\sqrt{R(a_1)}}\hat{B}_+
| \psi_0 \rangle.
\end{equation}

To identify the algebra we consider the commutation relations 
\begin{equation}
  \label{eq:17}
[ \hat{B}_- , \hat{B}_+ ] = R(a_0)
\end{equation}
\begin{equation}
  \label{eq:18}
[ \hat{B}_+ , R(a_0) ] = (R(a_1) - R(a_0)) \hat{B}_+,
\end{equation}
\begin{equation}
  \label{eq:19}
[ \hat{B}_+ , (R(a_1)-R(a_0))\hat{B}_+ ] = \{ (R(a_2) -
  R(a_1))-(R(a_1) - R(a_0))\} \hat{B}_+,
\end{equation}
and so on. In general there are an infinite number of such commutation
relations. If the quantities $R(a_n)$ satisfy certain relations one of
the commutators in this series may vanish. For such a situation the
commutation relations obtained up to that point plus their complex
conjugates form a Lie algebra with a finite number of elements. For
example if the condition 
\be
\label{19a}
(R(a_2) - R(a_1))-(R(a_1) - R(a_0)) = 0
\ee
is satisfied then the algebra is \cite{Balantekin:1997mg}  either
$SU(2)$ or $SU(1,1)$. Most of the exactly solvable one-dimensional 
problems in quantum mechanics can be described by this
algebra \cite{ialev}. It can be shown that this algebra also describes
for example both the bound and scattering states of the
P\"oschl-Teller potential \cite{Alhassid:zx} as well as associated
transfer matrices.

\subsection{Outlook on future applications}

Almost all exactly solvable one-dimensional potential problems
encountered in quantum mechanics textbooks are shape invariant where
the parameters are related by a translation \cite{Cooper:1994eh} 
\be
a_2 = a_1 + \eta . 
\ee
It should be emphasized that shape-invariance is not the most general
integrability condition one can write for such potentials as there 
are exactly solvable problems which are not shape invariant
\cite{Cooper:tz}. There is a second class of shape invariant
potentials where the parameters of the partner potentials are related
by a scaling \cite{Khare:gg,Barclay:1993kt}
\be
a_2 = q a_1 .
\ee
In this latter class, corresponding one-dimensional potentials are
defined implicitly, but explicit expressions are not given.

In searching for integrable models in two-dimensional statistical
mechanics a relationship was uncovered between those models,
three-dimensional Chern-Simons gauge theory and quantum groups
\cite{Witten:1989rw}. These models, being completely integrable, can
be written in a shape-invariant way \cite{bahaprep}, corresponding to
a shift in the parameters 
\be
a_2 = q a_1 + \eta .
\ee
The associated algebras are called up-down algebras
\cite{georgia}. These developments suggest that there may be
shape-invariant potentials where the parameters are related by 
linear-fractional transformations: 
\be
a_2 = ( q a_1 + \eta )/ ( a_1 + \eta')
\ee
This is a completely unexplored direction of research as nothing is
known about such integrable systems. Recall that the notation $a_1$,
$a_2$, etc. may represent not only single parameters, but also a set
of them. In general one may suggest to
simply relate these parameters by the transformation 
\be
\hat{T}(a_1) {\cal O}(a_1) \hat{T}^{-1}(a_1) = {\cal O}(a_2).
\ee
where $\hat{T}$ is an element of any group, not just of SL(2,R) as
suggested by the linear-fractional transformation and its limits that
were so far employed. What
kind of exactly solvable problems do we obtain? At the moment this is
an open question.     

The basic philosophy of this approach is to consider the parameters of
the Hamiltonians as auxiliary dynamical variables. This is reminiscent
of the path leading to the Interacting Boson Model
\cite{Armia:1976ky}. To describe the quadrupole collectivity in nuclei
one needs to consider a five-dimensional space. It is possible to
formulate this problem in terms of boson variables \cite{jolos},
however the problem is nonlinear written in terms of quadrupole
bosons. By considering a parameter of the problem (boson number) as an
additional degree of freedom, Interacting Boson Model introduced a
scalar boson as a dynamical variable. This has led to the subsequent
realization\cite {talmi1} of $s$ and $d$ bosons as pairs of 
nucleons coupled to the angular momentum $L=0$ and $L=2$

So far we talked about considering parameters of the shape-invariant
problem as auxiliary dynamical variables. One can imagine an
alternative approach of classifying some of the dynamical variables as
``parameters''. An example of this is provided by the supersymmetric
approach to the spherical Nilsson model of single particle states
\cite{Balantekin:1992qp}. The Nilsson Hamiltonian is given by 
\begin{equation}
  \label{eq:20}
H = \sum_i a^\dagger_i a_i - 2 k {\bf L . S} + k \nu {\bf L}^2 .
\end{equation}
Introducing the variable 
\begin{equation}
  \label{eq:21}
F^\dagger = \sum_i \sigma_i  a^\dagger_i
\end{equation}
one can show that the ``Hamiltonians'' 
\begin{equation}
  \label{eq:22}
H_1 = F^\dagger F = \sum_i a^\dagger_i a_i - {\bf \sigma . L}
\end{equation}
and
\begin{equation}
  \label{eq:23}
H_2 = F F^\dagger =  \sum_i a_i a^\dagger_i + {\bf \sigma . L}
\end{equation}
can be considered as supersymmetric partners of each other
\cite{Balantekin:1992qp}. The
shape-invariance condition of Eq. (\ref{eq:8}) can be written as 
\begin{equation}
  \label{eq:24a}
F F^\dagger = F^\dagger F + R, 
\end{equation}
where the remainder is
\begin{equation}
  \label{eq:25}
R= {\bf \sigma . L} - 3/4 ,
\end{equation}
i.e. in this example the radial variables are considered as the main
dynamical variables and the angular variables are considered as the
``parameters''. 

A number of applications of shape-invariance are available in the
literature. These include i) Quantum tunneling through supersymmetric
shape-invariant potentials \cite{Aleixo:1999cx}; ii) Study of neutrino
propagation through shape-invariant electron densities
\cite{Balantekin:1997jp}; iii) Investigation of coherent states for
shape-invariant potentials \cite{Balantekin:1998wj,Aleixo:2002sa}; and
iv) As attempts to devise exactly solvable coupled-channel problems, 
generalization of Jaynes-Cummings type Hamiltonians to shape-invariant
systems \cite{Aleixo:2000ub,Aleixo:2001jm}. In this article we focus
on the applications to many-body systems. 

\section{Many-Body Hamiltonians}

One can ask if these methods can be used to search for
exactly-solvable many-body systems. It has been shown that the concept
of supersymmetric shape-invariance can be utilized to derive the
energy spectrum of Calogero-Sutherland model \cite{Ghosh:1997yy}. 
Here we discuss an alternative approach and first write down multiple
commutators for a shape-invariant Hamiltonian 
\be
[\hat{H}, \hat{B}_+ ] = R(a_1) \hat{B}_+
\ee
\be
 [[\hat{H}, \hat{B}_+ ],\hat{B}_+ ] = (R(a_1) - R(a_2))\hat{B}_+^2
\ee
\be
[[[\hat{H}, \hat{B}_+ ],\hat{B}_+ ], \hat{B}_+ ] = (R(a_1) - 2R(a_2) +
  R(a_3))\hat{B}_+^3
\ee
\be
[[[[\hat{H}, \hat{B}_+ ],\hat{B}_+ ], \hat{B}_+ ], \hat{B}_+ ] =
(R(a_1) - 3 R(a_2) + 3 R(a_3) - R(a_4)) \hat{B}_+^4
\ee
and so on. We wish to address the possibility of defining an exactly
solvable problem through these commutation relations. We will consider
$\hat{B}_+$ as a raising operator. We assume that the Hamiltonian
$\hat{H}$ may or may not be in the form given by Eq. (\ref{eq:10}). 
We consider a generalized pairing problem with
\be
\label{37}
 \hat{B}_+ = \sum_j c_j S^+_j .
\ee
In Eq. (\ref{37}) the pair creation operator in a single-$j$ shell is
defined as  
\be
\label{singleshell}
 S^+_j = \sum_m \frac{1}{2} (-)^{j-m} a^{\dagger}_{j,m}
 a^{\dagger}_{j,- m},
\ee
where $a^{\dagger}_{j,m}$ is the particle creation operator. If we
assume that the shape-invariant Hamiltonian has only one- and two-body
terms the commutator $[[\hat{H}, \hat{B}_+ ],\hat{B}_+ ]$ will
only involve products of four creation
operators. Consequently the next nested commutator will vanish: 
\be
[[[\hat{H}, \hat{B}_+ ],\hat{B}_+ ], \hat{B}_+ ] = 0
\ee
Higher nested commutators will also vanish. This will place strong
constraints on $R(a_n)$, i.e. 
\be
 R(a_3) = - R(a_1) + 2R(a_2), 
\ee
\be
 R(a_4) = R(a_1) - 3 R(a_2) + 3 R(a_3)
\ee
and so on. Consequently we can immediately write the energy
eigenvalues and eigenstates of the Hamiltonian
\be
\hat{H} \hat{B}_+^n |\psi_0 \rangle = \left( n R(a_1) + \frac{1}{2} W
  n(n-1) \right) \hat{B}_+^n |\psi_0 \rangle ,
\ee
where 
\be
W = R(a_2) - R(a_1) .
\ee
A similar approach was first given by Talmi \cite{talmi}.

\section{Connection to Gaudin Algebras}

The pairing model with a constant two-body interaction was 
solved exactly by Richardson \cite{richardson}. In a parallel
development Gaudin developed an algebraic
approach to solve many-body spin Hamiltonians
\cite{gaudin1,gaudin2}. Here we will explore the relationship between
Gaudin's methods, algebraic methods developed to search for
quasi-exactly solvable models \cite{ush} and supersymmetric quantum
mechanics.  

Following the notation of Ref. [29] we 
consider the function defined as 
\be
\label{50}
\Psi (\lambda) = \prod_i^N \left( \lambda - \xi_i \right) e ^{- \int W
  d\lambda },
\ee
where $W (\lambda)$ is an arbitrary function of $\lambda$ and 
$\xi_i$ are numbers to be determined. 
Introducing the operators
\be
A = W + ip,  \,\,\,\,  A^\dagger = W - ip, 
\ee
it can be shown that the function defined in Eq. (\ref{50}) satisfies 
the equation   
\be
\label{50a}
A^\dagger A \,\,\Psi = \left[ 2 \sum_{i \neq j} \frac{1}{(\lambda -
    \xi_i)(\lambda - \xi_j)} - 2 \sum_i \frac{W(\lambda)}{(\lambda -
    \xi_i)} \right] \Psi .
\ee
Requiring the residue at $\xi_i$ to vanish yields the Bethe-ansatz 
conditions:
\be
\label{51}
W (\xi_i) =  \sum_{i \neq j} \frac{1}{\xi_i - \xi_j} .
\ee
Inserting Eq. (\ref{51}) into Eq. (\ref{50a}) we obtain
\be
\label{52}
A^\dagger A \,\,\Psi = 2 \sum_i \left(
  \frac{W(\lambda)-W(\xi_i)}{\lambda -
  \xi_i} \right) \,\, \Psi .
\ee                  
Provided that their superpotentials satisfy the condition given in 
Eq. (\ref{51}), factorized supersymmetric Hamiltonians satisfy
Eq. (\ref{52}). Note that the right side of Eq. (\ref{52}) in general
depends on $\lambda$, hence we cannot interpret the term that
multiplies the function $\Psi$ as an energy eigenvalue. However, 
for a number of 
limited cases (certain functions $W (\lambda)$ 
such as those that correspond to 
a harmonic oscillator) this $\lambda$ dependence drops out and
one can recover the standard expressions for the energy eigenvalues 
\cite{gernot}.   

The three generators of Gaudin's algebra ($J_0
(\lambda),J_{\pm}(\lambda)$) can be defined through the commutation
relations  
\be
[J_0 (\lambda), J_+ (\mu)] = - \frac{J_+ (\lambda) - J_+
  (\mu)}{\lambda - \mu} ,
\ee
\be
[J_-(\lambda), J_+ (\mu)] = - 2 \frac{J_0 (\lambda) - J_0
  (\mu)}{\lambda - \mu} ,
\ee
and 
\be
[ J_{\pm,0} (\lambda), J_{\pm,0} (\mu)] = 0 ,
\ee
where $\lambda$ is, in general, a continuous parameter. 
Gaudin studied the eigenstates of the ``Hamiltonian'' \cite{gaudin2} 
\be
\label{55}
H (\lambda) =  J_0 (\lambda) J_0 (\lambda) - \frac{1}{2} J_-(\lambda)
J_+(\lambda)  - \frac{1}{2} J_+(\lambda) J_-(\lambda)
\ee
If a state $\mid 0 \rangle$ which is annihilated by all $J_-
(\lambda)$ can be identified
\be
J_- (\lambda) \mid 0 \rangle = 0, 
\ee
then $W(\lambda)$ is introduced as the eigenvalue of $J_0 (\lambda)$
on that state: 
\be
J_0 (\lambda) \mid 0 \rangle = W (\lambda) \mid 0 \rangle .
\ee
One can then find the eigenvalues and eigenstates of the
``Hamiltonian'' of Eq. (\ref{55}): 
\be
\label{56}
H (\lambda) \mid \Phi \rangle = E (\lambda) \mid \Phi \rangle ,
\ee
where the eigenstates are 
\be
\mid \Phi \rangle = J_+ (\xi_N) J_+ (\xi_{N-1}) \cdots J_+ (\xi_1)
\mid 0 \rangle ,
\ee
and the eigenvalues are 
\be
E (\lambda) = W^2 (\lambda) + W' (\lambda) + 2 \sum_i \left(
  \frac{W(\lambda)-W(\xi_i)}{\lambda -
  \xi_i} \right) \,\, .
\ee
In deriving the above equations the conditions 
\be
W (\xi_i) = - \sum_{i \neq j} \frac{1}{\xi_i - \xi_j}, \,\, i,j = 1, 
\cdots, N  
\ee
were assumed to be fulfilled. 

The strategy of using Richardson-Gaudin methods to deal with many-body
problems were employed by a number of authors 
\cite{Pan:1997rw,Dukelsky:2001bc,Dukelsky:2001fe,Pan:2002yq}. 
Clearly there is a mapping between the solutions of the Gaudin problem
(Eq. (\ref{56})) and those of the factorized supersymmetric 
Hamiltonians (Eq. (\ref{52})). One may ask if this correspondence can
be exploited to study pairing and related problems. 

The pair creation operator\cite{Balantekin:1992qp} in a 
single-$j$ shell is defined in 
Eq. (\ref{singleshell}), 
\be
 S^+_j = \sum_m \frac{1}{2} (-)^{j-m} a^{\dagger}_{j,m}
 a^{\dagger}_{j,- m} ,
\ee   
its Hermitian conjugate, and the number operator span an $SU(2)$
algebra (the so-called quasi-spin algebra). One can obtain a Gaudin
algebra from the quasi-spin algebra by defining 
\be               
S_+ (\lambda) = \sum_j  \frac{S^+_j}{\lambda - \epsilon_i},
\ee                 
(and similar formulas for the other elements). This realization of the
Gaudin algebra can be very useful in many-body systems. As a simple
example we consider a system with $s$ and $p$ bosons and define
three operators that satisfy Gaudin's commutation relations 
\be
B_+(\lambda) = \frac{1}{2} \left[ \frac{s^\dagger s^\dagger}{\lambda -
    \alpha_s} + \frac{(p^\dagger \cdot p^\dagger)}{\lambda -
    \alpha_p} \right]
\ee
\be
B_-(\lambda) = [B_+(\lambda)]^\dagger ,
\ee
and
\be
B_0(\lambda) =  \frac{1}{2} \left[ \frac{\hat{n}_s}{\lambda -
    \alpha_s} + \frac{\hat{n}_p + 3/2}{\lambda -
    \alpha_p} \right].
\ee
It is easy to show that as $\alpha_s \rightarrow \alpha_p$ the 
quantity $B_+(\lambda)B_-(\lambda)$ reduces to 
\be
\frac{1}{\lambda - \alpha_p} \hat{P}_4
\ee
where $\hat{P}_4$ is the O(4) pairing operator. One can then study a
Gaudin-type Hamiltonian which generalizes this operator
\be
H (\lambda) = B_0(\lambda)B_0(\lambda) - \frac{1}{2}
B_-(\lambda)B_+(\lambda) B_+(\lambda)B_-(\lambda).
\ee
Following steps above one can show that this Hamiltonian is associated
with the one-dimensional potential
\be
V(x) = \frac{1}{2} \left( \frac{1}{x - \alpha_s} + \frac{1}{x -
    \alpha_p} \right)^2 .
\ee 
Similar ideas could conceivably be useful in dealing with other
many-body systems.

\vspace*{-2pt}

\section*{Acknowledgments}
I thank F. Iachello for his support, encouragement, and friendship
over many years. 
This work was supported in part by the U.S. National Science
Foundation Grants No. PHY-0136261 and PHY-0070161. 

\end{document}